\documentclass[a4paper]{spie}  

\usepackage[]{graphicx}
\usepackage[]{epsfig}
\newcommand{\xmm}{{XMM-{\em Newton} }}

\title{Processing challenges in the XMM-Newton slew survey}

\author{Richard D. Saxton\supit{a}, Bruno Altieri\supit{a},
Andrew M. Read\supit{b}, Michael J. Freyberg\supit{c},
M. Pilar Esquej\supit{a} and Diego Bermejo\supit{a}
\skiplinehalf
\supit{a}XMM-SOC, ESAC, Villafranca del Castillo, Apartado 50727,
28080 Madrid, Spain; \\
\supit{b}Department of Physics and Astronomy, University of Leicester, Leicester Le17RH, England; \\
\supit{c}Max-Planck-Institut fuer Extraterrestrische Physik, PO Box 1312, 85741 Garching, Germany. \\
}

\authorinfo{Further author information: (Send correspondence to R.D.S.)\\R.D.S.: E-mail: Richard.Saxton@sciops.esa.int, Telephone: +34 91 8131306\\  B.A.: E-mail: Bruno.Altieri@sciops.esa.int, Telephone: +34 91 8131340}

\begin{document} 
  \maketitle 

\begin{abstract}
The great collecting area of the mirrors coupled with the high
quantum efficiency of the EPIC detectors have made XMM-Newton
the most sensitive X-ray observatory flown to date. This is
particularly evident during slew
exposures which, while giving only 15 seconds of on-source time, actually
constitute a 2-10 keV survey ten times deeper than
current "all-sky" catalogues. Here we report on progress towards
making a catalogue of slew detections constructed from the full,
0.2-12 keV energy band and discuss the challenges associated
with processing the slew data. The fast (90 degrees per hour) slew
speed results in images which are smeared, by different amounts
depending on the readout mode, effectively changing the form of the
point spread function. The extremely low background in slew images changes
the optimum source searching criteria such that searching a single image using
the full energy band is seen to be more sensitive than splitting the
data into discrete energy bands. False detections due to
optical loading by bright stars, the wings of the PSF in very bright sources
 and single-frame detector flashes are
considered and techniques for identifying and removing these spurious
sources from the final catalogue are outlined.
Finally, the attitude reconstruction of the satellite during the slewing
manoeuver is complex. We discuss the implications of this on the 
positional accuracy of the catalogue.
\end{abstract}



\section{INTRODUCTION}
\label{sect:intro}  

\xmm\cite{Jansen01} performs slewing manoeuvers between observation targets
with the EPIC cameras open and the other instruments closed. Both
EPIC-pn\cite{Struder01} and EPIC-MOS\cite{Turner01} are operated during slews with
the Medium filter in place and the observing mode set to that of the
previous observation.

The satellite moves between targets by performing 
an open-loop slew along the roll and pitch axes
and a closed-loop slew, where  measurements from the star tracker are used in addition to the
Sun--sensor measurements to provide a controlled slew about all three axes, to correct for residual errors in the
long open-loop phase. The open-loop slew is performed at a steady rate
of about 90 degrees per hour and it is data from this phase which may be
used to give a uniform survey of the X-ray sky.

\begin{figure}
\begin{center}
\begin{tabular}{c}
\includegraphics[height=6cm]{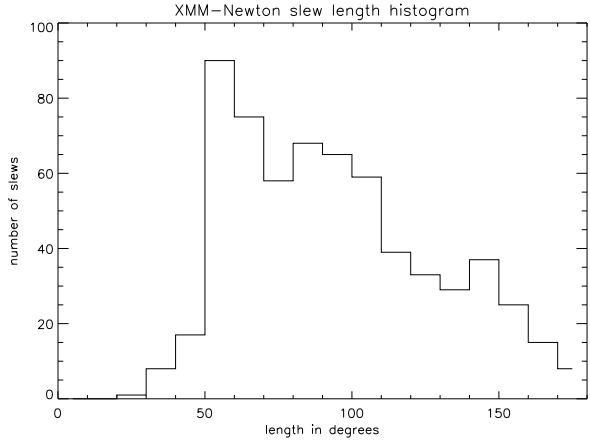}
\end{tabular}
\end{center}
\caption[Slew length histogram]
{ \label{fig:slewlngth}
A histogram of the distance across the sky covered by individual slews.}
\end{figure}

Slew Data Files (SDF) have been stored in the \xmm Science Archive (XSA) 
from revolution 314 and there are currently 465 SDFs stored
with a mean slew length of 86 degrees (Fig.~\ref{fig:slewlngth} ).
Not all of these data are scientifically useful and the sky coverage will
be discussed in Sect.~\ref{sect:obs}. 

The data are being used to perform three independent surveys, a soft band
(0.2--2 keV) X-ray survey with strong parallels to the ROSAT all--sky survey
\cite{Voges99}(RASS), a hard band (2--12 keV) survey and an \xmm full-band
(0.2--12 keV) survey.

Theoretically the good point spread function of the X-ray telescopes
\cite{Aschenbach00} should allow source positions to be determined to an 
accuracy of better than 6 arcseconds, similar to that found for faint 
objects in the 1XMM catalogue of serendipitous sources detected in pointed
observations (\footnote{The First XMM-Newton Serendipitous Source Catalogue,  XMM-Newton Survey Science Centre (SSC), 2003.}). Any errors in the attitude reconstruction 
for the slew could seriously degrade this performance and a major technical
challenge of the data processing is to achieve the nominal accuracy. We
address this issue in Sect.~\ref{sect:attitude}.

\section{Observations and data analysis}
\label{sect:obs}  

\begin{figure}
\begin{center}
\begin{tabular}{c}
\rotatebox{-90}{\includegraphics[height=8cm]{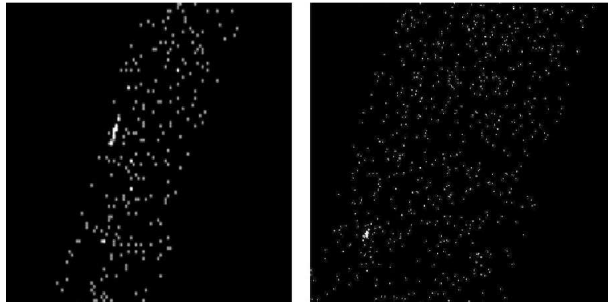}}
\end{tabular}
\end{center}
\caption[The appearance of slew sources]
{ \label{fig:psf}
The appearance of a source in a MOS (left) and EPIC-pn (right) slew image. Note the extension in the slew travel direction in the MOS image due to integration over a frame time of 2.6s.}
\end{figure}

The appearance of a source in the slew depends on the frame time 
of the observing mode as photons can only be positioned in space to 
an accuracy of one frame. This has major implications for MOS, where the
relatively long frame time of 2.6 seconds, spreads out a source into a 4 arcminute
long streak (Fig.~\ref{fig:psf}). EPIC-pn has a much faster readout and
source extension in the slew direction is less than 18 arcseconds
in all observing modes. 
The relatively large source profile in the MOS cameras
and their lower effective area make the EPIC-pn a much superior instrument
for performing a slew survey. For this reason only pn data are being used
in the data analysis.

\begin{figure}
\begin{center}
\begin{tabular}{c}
\rotatebox{-90}{\includegraphics[height=7cm]{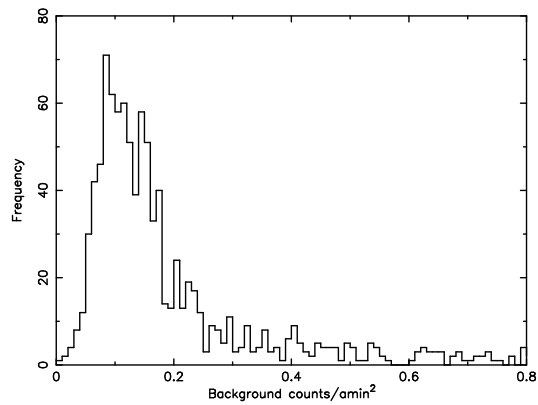}}
\end{tabular}
\end{center}
\caption[Background levels]
{ \label{fig:bgnd}
A histogram of the total band (0.2-12 keV) EPIC-pn background levels in units
of counts/arcminute$^{2}$.}
\end{figure}

Sources pass through the field of view of EPIC-pn in about 15 seconds.
This low exposure time leads to a generally very low background 
of average 0.1 c/arcminute$^{2}$
in normal conditions. However, some slews taken at times of enhanced
solar activity do exhibit higher background (Fig.~\ref{fig:bgnd}) and can
give rise to a large number of spurious sources.
For each slew we have computed average band rates in 6 energy bands
to characterize the general rate level.
In this first processing slews with high background that had an average
count rate exceeding 5.5 c/s in the $7.5-12$\,keV band are
being discarded.
In later processings we will also use low-background periods
of contaminated slews using time selections, which will yield
another $\sim 10\%$ exposure time.

\begin{figure}
\begin{center}
\begin{tabular}{c}
\rotatebox{-90}{\includegraphics[height=7cm]{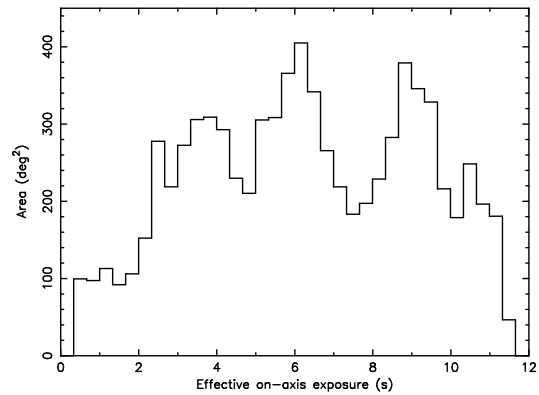}}
\end{tabular}
\end{center}
\caption[Sky area as a function of effective exposure time]
{ \label{fig:expos}
A histogram of the total band (0.2-12 keV) EPIC-pn sky coverage as a
function of effective on-axis exposure time. The histogram has been normalised
to a total of 8,000 square degrees.}
\end{figure}

A total of 605 slew datasets have been processed and stored in the XSA.
Of these 424 have been made with pn in the useful FF (295), eFF (88)
and LW (41) modes respectively. After removal of the high background
slews we are left with 312 slews, with a mean sky area of the useable
part of the data of $\sim25$ square degrees,  giving a total sky coverage 
of some 8,000
square degrees, ignoring overlaps. The sky coverage is uniform but
subject to the vignetting function such that sources passing directly 
through the centre of the detector receive an equivalent of 11 seconds
of on-axis exposure while sources further from the centre receive less.
The mean equivalent on-axis exposure time over the sky is 6.3 seconds
and the sky area covered as a function of exposure time is shown in 
Figure~\ref{fig:expos}.

Events are recorded initially in RAW or detector coordinates and have to 
be transformed, using the satellite attitude history, into sky coordinates.
The tangential plane geometry commonly used to define a coordinate grid for
flat images is only valid for distances of 1--2 degrees from a reference
position, usually placed at the centre of the image. To avoid this 
limitation, slew datasets are divided into roughly one square degree
event files, attitude corrected and then converted into images.
This relies on the attitude history of the satellite being accurately known
during the slew; a point which is addressed in section~\ref{sect:attitude} .

\subsection{Instrumental aspects}
\label{sect:instasp}  

The \xmm {\em Slew Data Files} (SDFs) for EPIC-pn
were processed using the {\tt epchain} package of the
public {\tt xmmsas-6.1} plus a small modification for
the {\tt oal} library.
For diagnostic reasons a few parameters were set to
non-default values (e.g., keeping also events below
150\,eV).

For the {\em Slew Survey} catalogue 
we selected only EPIC-pn exposures
performed in {\em Full Frame} (FF),
{\em Extended Full Frame} (eFF), and {\em Large Window}
(LW) modes, i.e.\ modes where all 12 CCDs are 
integrating (in LW mode only half of each CCD). 
The corresponding cycle times are
73.36\,ms, 199.19\,ms, and 47.66\,ms, which converts
to a scanned distance of 6.6 arcseconds, 17.9 arcseconds, and
4.3 arcseconds per cycle time, respectively.
In the {\em Small Window} mode only the
central CCD is operated and a window of $64\times64$
pixels is read out, i.e.\ only about 1/3 of a CCD.
In the fast modes, {\em Timing} and {\em Burst},
only 1-dimensional spatial information 
for the central CCD is available
and thus these modes are not very well suited for
source detection. Therefore for these three modes the
{\em Closed} filter position will be used in the future instead of
the {\em Medium} position, to utilize this unsuable
exposure time for calibration purposes.

\subsection{Source search procedure}
\label{sect:search}  

Pilot studies were performed to investigate  the optimum
processing and source-search strategies. By making small changes 
to the \xmm standard analysis software
(SAS) we have been able to successfully create and use correct exposure maps
in the source searching.  These produced
no unusual effects, although uneven (and heightened) slew exposure is
observed at the end of slews (the 'closed-loop' phase). We
tested a number of source-searching techniques and found that the
optimum source-searching strategy was usage of a semi-standard
`eboxdetect (local) + esplinemap + eboxdetect (map) + emldetect'
method, tuned to $\sim$zero background, and performed on a single
image containing just the single events (pattern=0) in the
0.2$-$0.5\,keV band, plus single and double events (pattern=0$-$4) in
the 0.5$-$12.0\,keV band. This is similar to the technique used for
producing the RASS catalogue \cite{Cruddace88} and resulted in the 
largest numbers of
detected sources, whilst minimising the number of spurious sources
due to detector anomalies (usually caused by non-single, very soft
($<$0.5\,keV) events). The source density was found to be $\approx$0.5
sources per square degree to an emldetect detection likelihood threshold
(DET\_ML) of 10 (approx 3.9$\sigma$).

In the current and on-going slew pipeline, images and exposure maps
have been created and source-searched. This is being done in 3
separate energy bands: full band (0.2$-$0.5\,keV [pattern=0] +
0.5$-$12.0\,keV [pattern=0$-$4]), soft band (0.2$-$0.5\,keV
[pattern=0] + 0.5$-$2.0\,keV [pattern=0$-$4]), and hard band
(2.0$-$12.0\,keV [pattern=0$-$4]). In this processing we are now 
recording detections down
to an {\it emldetect} detection likelihood threshold of 8 (approx
3.4$\sigma$), and detecting $\approx$0.7 sources per square degree.

\section{Spurious sources}
\label{sect:spurious}  

Systematic effects exist with the instrument and detection software which 
lead to a number of spurious detections. The three principle causes are 
outlined below.

\subsection{Optical loading}
EPIC-pn slew exposures are possibly affected by optical loading contamination.
This effect is due to several optical photons (each creating a 3.65 eV charge) piling-up
above the low-energy threshold of 20 ADUs and creating fake X-ray counts.
On pointed observation this effect is removed by offset maps acquired at the start
of each exposure and subtracted on-board, but this is not the case for slew exposures,
where the offset map of the previous (pointed) exposure is applied.
As a consequence, very bright stars could be affected by optical loading in the XMM
slew survey.

Based on the measured optical transmission of the Medium filter and theoretical
considerations, optical loading is expected for stars brighter than magnitude V=3.7,
where more than 5 counts would be due optical photons.

\begin{figure}
\begin{center}
\begin{tabular}{c}
\rotatebox{0}{\includegraphics[height=5cm]{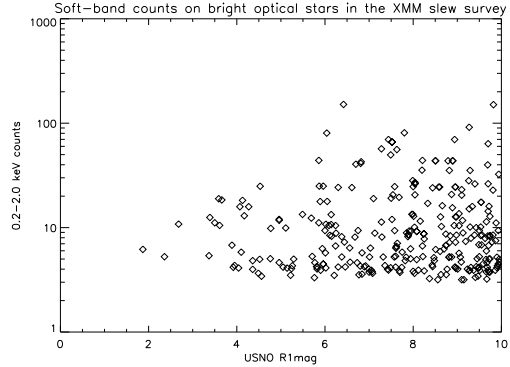}}
\end{tabular}
\end{center}
\caption[]
{ \label{fig:opload}
A plot of counts detected in the soft band against R magnitude for
counterparts taken from the USNO catalogue.}
\end{figure}

The optical loading has been assessed using bright USNO stars detected in the
slew survey. Figure~\ref{fig:opload} shows soft band slew counts plotted against their R magnitude.
Stars fainter than R=4 are not affected by optical loading, as no correlation is found
between their count rate and magnitude.
Stars brighter than R=4 could possibly be affected by optical loading counts although it is not yet clear to what extent. Some evidence shows that it would play
only a minor role for stars down to R=2. Two V=2.7 stars have been detected so far with less than 10 counts, much
less than expected from optical loading only, so optical loading might not be an issue
at all. 

\subsection{Detector flashes}

\begin{figure}
\begin{center}
\begin{tabular}{c}
\rotatebox{-90}{\includegraphics[height=7cm]{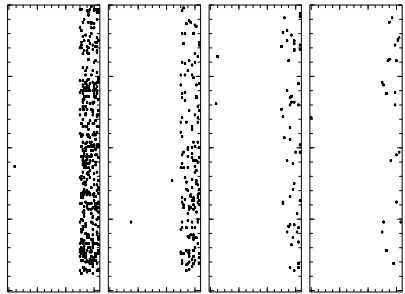}}
\end{tabular}
\end{center}
\caption[]
{ \label{fig:slew_flash_0365}
Four consecutive EPIC-pn frames showing a large number of
(low-energy) events distributed along neighbouring columns;
these features can easily be discriminated from astronomical sources
in detector and time space (but are more difficult to distinguish 
once projected onto the sky).
}
\end{figure}

We have created lightcurves with short time bin size ($<1$\,s)
in the softest channels to identify short-duration CCD flashes
that occur only for $< 200$\,ms in several adjacent CCD columns
with a very soft spectrum. Projected onto the sky these can lead
to spurious sources. Figure~\ref{fig:slew_flash_0365} shows
4 consecutive readout frames containing one of these flashes.

These effects are minimised by only using single-pixel (pattern 0)
events for photon energies less than 500 eV. Nevertheless, some
flashes may be manifest in slew images and so sanity checks of data 
and detector performance are made on the
basis of diagnostic images and lightcurves of each individual CCD.

\subsection{The wings of very bright sources}

\begin{figure}
\begin{center}
\begin{tabular}{c}
\rotatebox{-90}{\includegraphics[height=7cm]{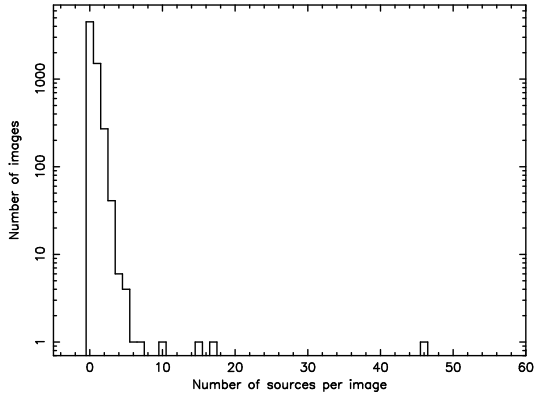}}
\end{tabular}
\end{center}
\caption[Histogram of source numbers per image]
{ \label{fig:numcount}
A histogram of the number of sources found in a single one degree image.
}
\end{figure}

It was noticed in the creation of the 1XMM serendipitous source catalogue
that, due to the imperfect modelling of the PSF, a halo of false detections
is often seen around bright sources. The same effect is seen in slew
exposures but due to the reduced exposure time is only important for 
very bright sources $\gg 10$ c/s. In addition large extended sources
often result in multiple detections of the same object. It is fairly easy to identify 
occurrences by searching for images with a large number of sources.
A histogram of source counts (Fig.~\ref{fig:numcount}) shows several outliers
from the main distribution including one image containing 46 sources;
which is actually due to Puppis-A (see Fig~\ref{fig:puppis}).

\section{Attitude reconstruction and positional accuracy}
\label{sect:attitude}

This section describes the issue of attitude reconstruction in slew 
observations,
which is crucial in the determination of source coordinates.
After showing how the 
attitude reconstruction is generally performed, we will concentrate on that 
of the open-loop slews which have been used in this survey.

The attitude information of the XMM-Newton satellite is provided by 
the Attitude and Orbit Control Subsystem (AOCS). A star tracker co-aligned 
with the telescopes allows up to a maximum of five stars to be continuously 
tracked giving accurate star position data every 0.5 seconds, which
operates in addition to the Sun 
sensor that provides a precise Sun-line determination. Such information is 
processed resulting in an absolute accuracy of the reconstructed astrometry 
of typically 1 arcsecond. For the open-loop slews, large slews outside the 
star-tracker field of view of 3 x 4 degrees, the on-board software generates a 
three axis momentum reference profile and a two-axis (roll and pitch) 
Sun-sensor profile, both based on the ground slew telecommanding. During  
slew manoeuvring a momentum correction is superimposed onto the reference 
momentum profile and, as there are no absolute measurements for the yaw axis, 
a residual yaw attitude error exists at the end of each slew that may be 
corrected in the final closed-loop slew.

So far, two types of attitude data can be used as the primary 
source of spacecraft positioning during event files processing. 
They are the Raw Attitude File (RAF) and the Attitude History File (AHF). 
For pointed observations, the RAF provides the attitude information at the maximum possible rate, 
with one entry every 0.5 seconds while the AHF is a smoothed and filtered version 
of the RAF, with times rounded to the nearest second. In slew datasets the
RAF stores attitude information every 40--60 seconds while the AHF
contains identical positions with timing information in integer seconds.
The user can select which one to use for data processing by 
setting an environment variable. 

In a pilot study where the AHF was used for attitude reconstruction, 
source detection was performed and their correlations with ROSAT and 2MASS 
catalogues indicated a slew relative pointing accuracy of $\sim10$ arcseconds, 
enough for a good optical follow up of the sources. However, an absolute 
accuracy of 0-60 arcseconds (30 arcsecond mean) was obtained in the slew direction, 
resulting in a thin, slew-oriented error ellipse around each source. 
This error appears to be consistent with the error introduced by the
quantisation of the time to 1 second in the attitude file and leads us to change the processing software
as a better accuracy should be obtained. Investigating further the errors, the 
RAF was used to compute the astrometry for some observations as a test. 
In this case, an offset of $\sim1$ arcminute from the ROSAT positions was found, 
but with a smaller scatter compared with the positions returned by the AHF
processing. The consistency of these offsets suggested that 
they could be due to a timing issue. This has been confirmed by flight dynamics
who stated that the tracking of up to five stars, mentioned above, produces 
a delay from the CCD exposure to data availability of approximately 0.75 seconds.
Subtracting directly this 0.75 seconds from every entry in the 
RAF we obtain an optimal attitude file for the processing.

\begin{figure}
\begin{center}
\begin{tabular}{c}
\rotatebox{90}{\includegraphics[height=9cm]{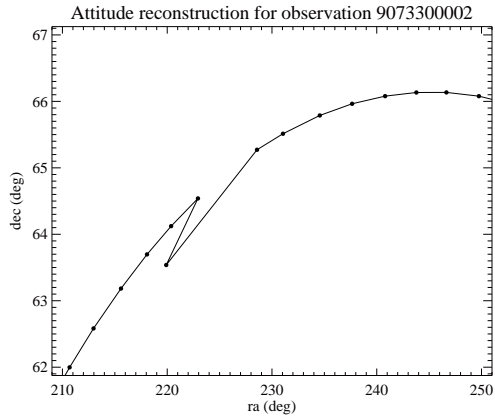}}
\end{tabular}
\end{center}
\caption[Rev 0733 peak]
{ \label{fig:peak}
A zoom into the problematic region of the attitude file in the slew 9073300002 before
    removing the bad RAF point. The points show the generated attitude information and the line shows the interpolation between the points. After the correction such a line becomes smooth.}
\end{figure}

\begin{figure}
\begin{center}
\begin{tabular}{c}
\rotatebox{90}{\includegraphics[height=9cm]{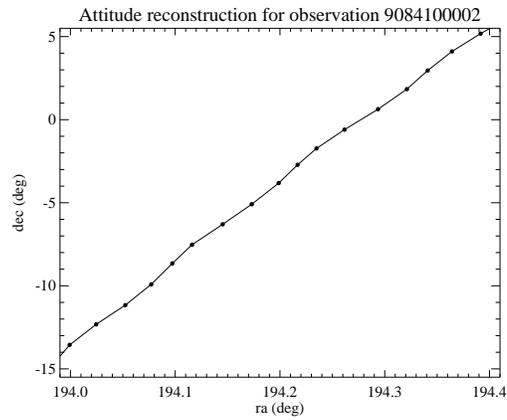}}
\end{tabular}
\end{center}
\caption[Rev 0841 turbulence]
{ \label{fig:turbulence}
A plot showing non-smooth, or turbulent, attitude reconstruction in
the revolution 0841 attitude file.}
\end{figure}

Other issues affecting the astrometry performance appeared after a careful 
visual examination of the RAF files, where two types of peculiarities 
appeared in some of the slews both affecting a localised region or the totality
of the slew. This means that if a source lies in a
problematic region its position has not been correctly generated.
On the one hand, 5 observations presented sharp discontinuities revealing the 
existence of single bad RAF points that have to be determined and removed from
the attitude file before performing the source searching. As an example
a source in the slew 9073300002 was discovered to have a closest ROSAT counterpart at 
a distance of 8 arcminutes. Investigation showed that the source was observed at a time 
coincident  with a large error in the attitude file (Fig.~\ref{fig:peak}). After the bad RAF point 
was removed the recalculated source position lies at 11 arcseconds from the ROSAT position.
On the other hand the attitude reconstruction of some slews appeared not smooth but 
turbulent in 7 observations (Fig.~\ref{fig:turbulence}) and this case is still under investigation.

\begin{figure}
\begin{center}
\begin{tabular}{c}
\hbox{
\epsfig{file=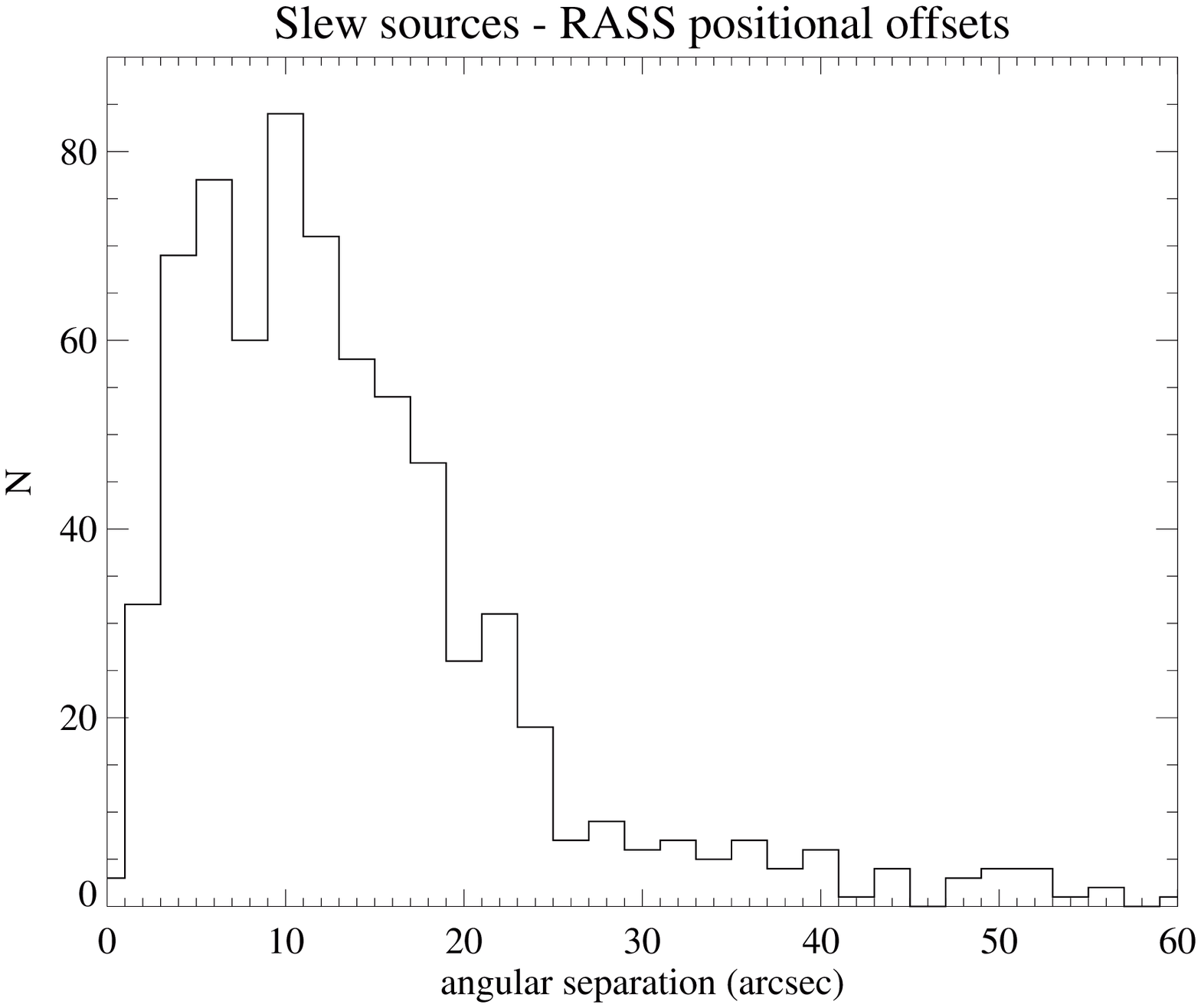,angle=90,width=8cm}
\epsfig{file=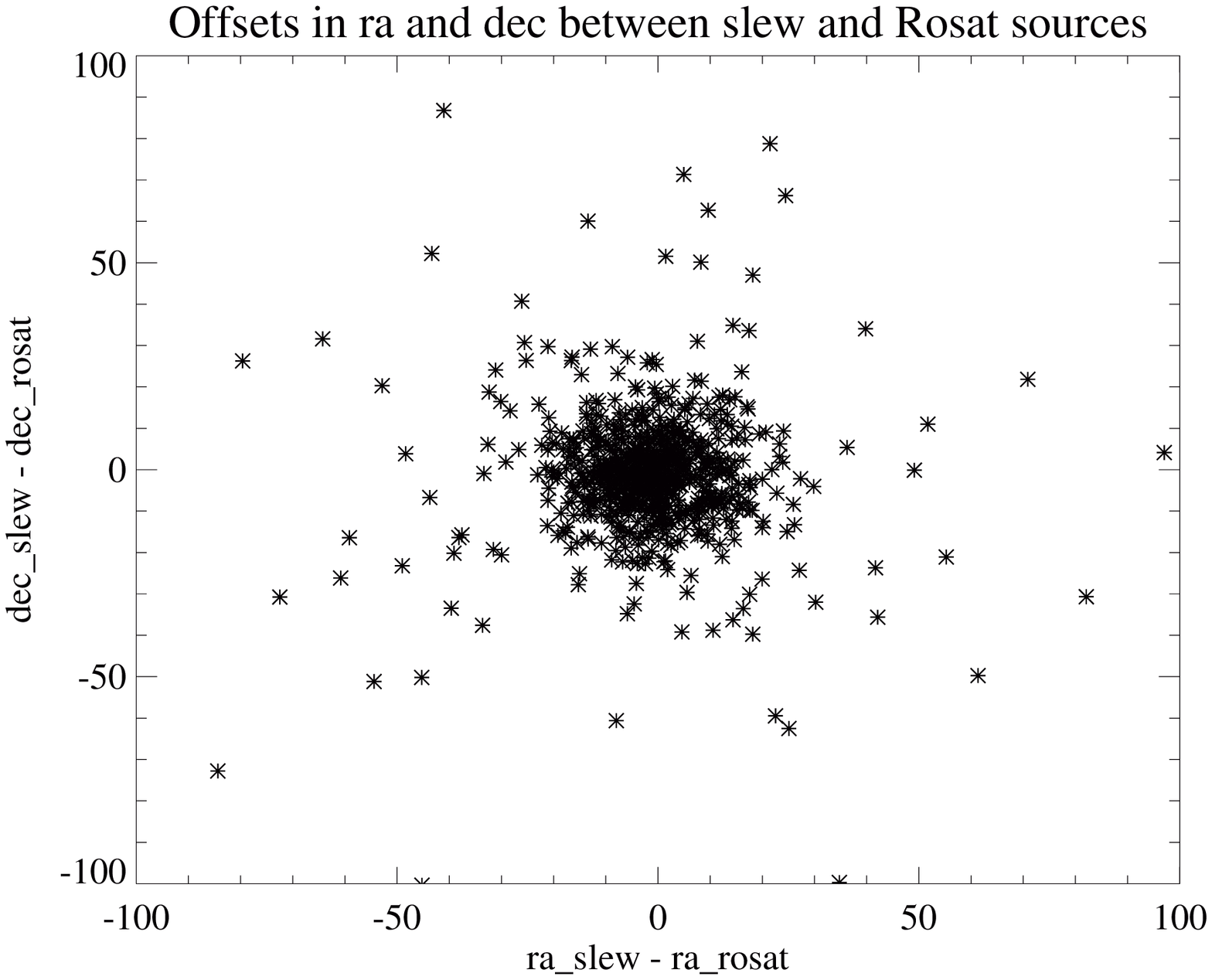,angle=90,width=8cm}
} 
\end{tabular}
\end{center}
\caption[RASS comparison]
{ \label{fig:ROSAThist}
A comparison of slew source positions with those from
the RASS catalogue. 68\% of the sources lie within 16 arcseconds.
The left panel shows a histogram of the offset magnitude while the
right panel gives the absolute offset in ra and dec  of the slew source
from the ROSAT position.}
\end{figure}

\begin{figure}
\begin{center}
\begin{tabular}{c}
\rotatebox{90}{\includegraphics[height=9cm]{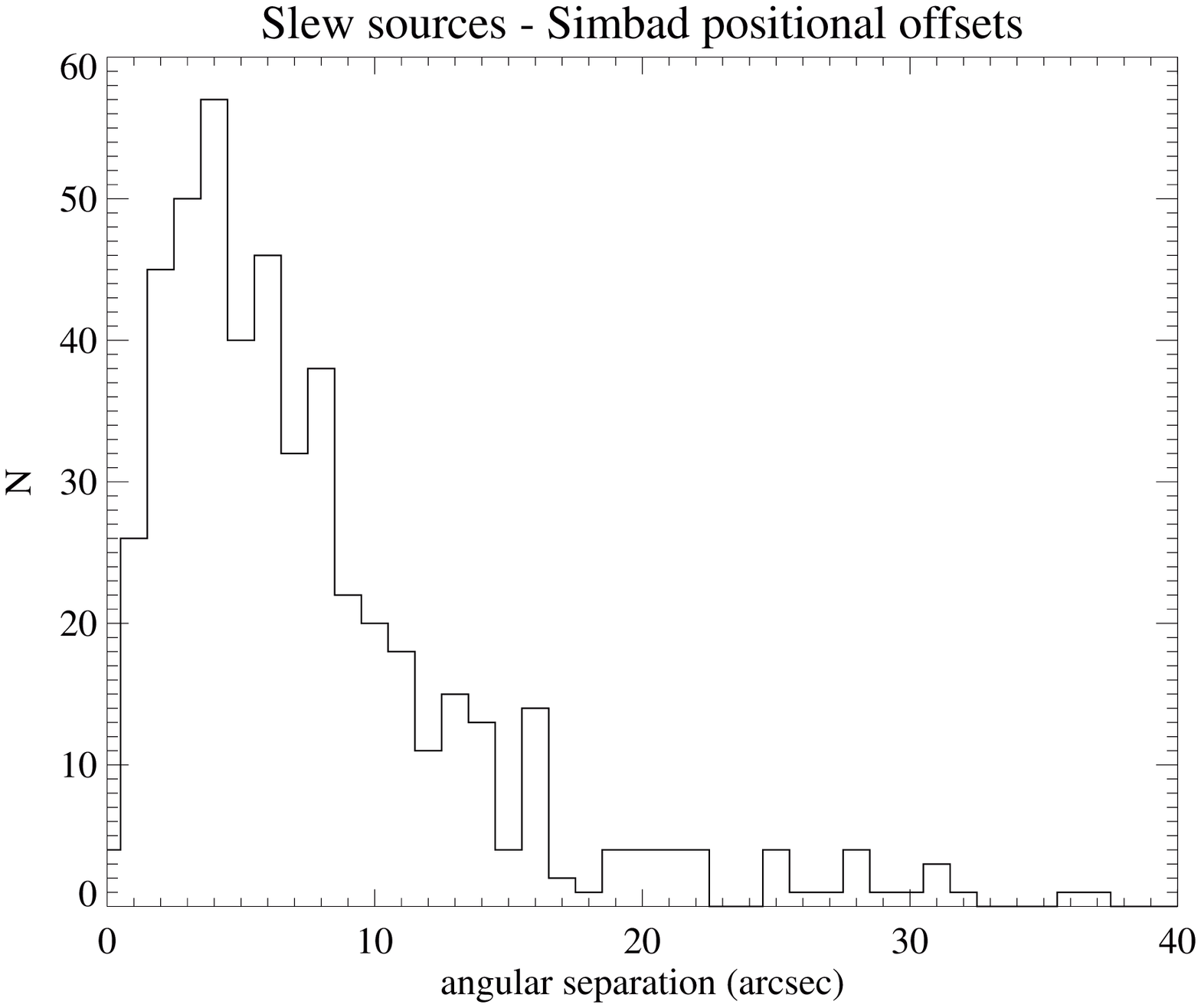}}
\end{tabular}
\end{center}
\caption[Simbad cross-correlation]
{ \label{fig:Simbadhist}
A histogram of the distribution of the angular separation in
    arcsecond derived from the correlation of the slew sources with the Simbad database
    within a 40 arcsecond offset. 68\% of the matches lie within 8 arcseconds.}
\end{figure}

Slew observations have been reprocessed using corrected RAFs and a
subsample of 1260 non-extended sources (defined as having an extent parameter
$<2$ from the emldetect source fitting) with DET\_ML$>10$, have been correlated with
several catalogues within a 60 arcsecond offset.
The correlation with the RASS reveals that $\sim60$\% of the slew sources
have an X-ray counterpart of which 68\% (90\%) lie within 16 (31) arcseconds 
(Fig.~\ref{fig:ROSAThist}). 
This gives confidence that the
majority of slews have well reconstructed attitude. Tests on some of
the outliers show that closer matches are sometimes available using
ROSAT pointed data from the 2RXP and 1RXH catalogues.
To form a sample of catalogues with highly accurate positions
but which minimise the number of false matches, we used the Astrophysical Virtual Observatory (AVO)
to correlate the slew positions against non-X ray SIMBAD catalogues. This 
gave 508 matches of which 68\% (90\%) were contained within 8 (17) arcseconds (Fig.~\ref{fig:Simbadhist}),
showing that the positional accuracy of the slew is not much worse than the 
observed limit for low significance \xmm sources.

\section{Results}
\label{sect:results}  

To date 138 datasets have been processed giving 2370 sources with
DET\_ML$>8$ (1600 with DET\_ML$>10$) in the total band and 440 in 
the hard X-ray band (220 with DET\_ML$>10$). A small pilot study 
visualising all DET\_ML$>10$ sources from ten slews, showed that
apart from the problems detailed in section~\ref{sect:spurious}, 
sources appeared
to be real. More sophisticated statistical tests or simulations 
will have to be applied 
to calculate the fraction of sources with $DET\_ML$
between 8 and 10 which are due 
to background fluctuations.

A great variety of sources have been detected, including stars,
galaxies, both interacting and normal, AGN, clusters of galaxies and SNR
plus extremely bright Low-Mass-X-ray 
Binaries (LMXB),
with several hundred c/s. These are bright enough to give a
useful spectrum although they suffer seriously from photon pile-up.
As we are essentially
performing three separate surveys, we have immediate access
to hardness ratios for many of the detected sources, and a large
variation in source hardness is seen. About one percent of sources are
detected in more than one slew, yielding short to medium term
(days to months) variability information. One
source, so far detected in three separate slews, appears to have
varied in flux by a factor of $\sim$2.

\begin{figure}
\begin{center}
\begin{tabular}{c}
\rotatebox{90}{\includegraphics[height=9cm]{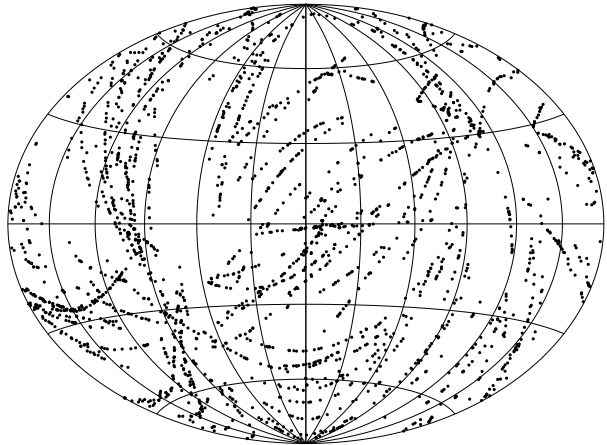}}
\end{tabular}
\end{center}
\caption[sky source plot]
{ \label{fig:skysource}
An Aitoff projection of the distribution of all slew sources on the sky in
galactic coordinates.}
\end{figure}

\begin{figure}
\begin{center}
\begin{tabular}{c}
\rotatebox{-90}{\includegraphics[height=8cm]{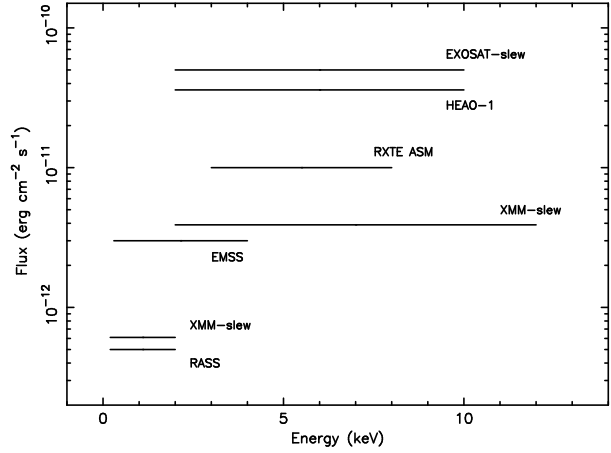}}
\end{tabular}
\end{center}
\caption[Flux limit plot]
{ \label{fig:fluxlim}
The flux limits of X-ray large area surveys. The \xmm limits have been 
calculated for a DET\_ML=10 source, with an absorbed power-law spectrum
of slope 1.7 and N$_{H}$=3.0$\times 10^{20}$$cm^{-2}$, passing through 
the centre of the field of view.}
\end{figure}

Figure~\ref{fig:skysource}  shows the distribution of sources over the sky
indicating the paths of slews processed so far.
The flux limits for the three surveys are compared with those
of other missions in Fig~\ref{fig:fluxlim}. At a DET\_ML of 10(8) sources are
detected to a flux limit of 6(4.5)$\times10^{-13}$ergs s$^{-1}$ cm$^{-2}$
in the soft band and 4(3)$\times10^{-12}$ergs s$^{-1}$ cm$^{-2}$
in the hard band. The mean flux limit over the whole
survey, taking into account the variable effective on-axis exposure
time, is 60\% higher than these values.

\subsection{Global correlations with the ROSAT all-sky survey}

\begin{figure}
\begin{center}
\begin{tabular}{c}
\rotatebox{90}{\includegraphics[height=9cm]{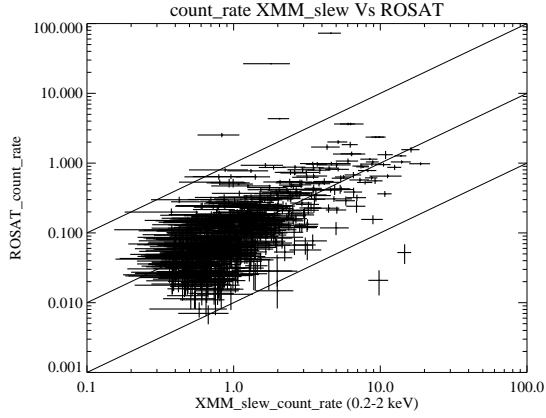}}
\end{tabular}
\end{center}
\caption[Puppis-A]
{ \label{fig:cratio}
A comparison of the count rates in the XMM soft band and the RASS.
The central line represents the mean ratio of $\sim10$ and the outer
lines variability by a factor of ten.}
\end{figure}

There is a strong overlap between the soft XMM slew catalogue and the
ROSAT all-sky survey (RASS) which is limited by statistics at the 
faint flux end of
each survey and by intrinsic source variability. Of the non-extended, 
DET\_ML$>10$ sources in the soft band slew survey 64\% have counterparts within 1
arcminute in the RASS. The fraction drops to 53\% for hard band slew sources.
A comparison of the two surveys reveals a mean count rate ratio of $\sim10$ 
(Fig.~\ref{fig:cratio}),
with one percent of the sources detected in both surveys showing variability 
by a
factor in excess of 10. Many more variable sources will be 
identified by performing an upper limits analysis on data from both surveys.
The combination of these surveys will enable the long term X-ray variability
of several thousand sources to be studied over a baseline of 10--15 years.
The high variability sources so far identified are formed from Blazars, low-mass X-ray binaries, eclipsing binaries and Seyfert I galaxies.

\subsection{Extended sources}

\begin{figure}
\begin{center}
\begin{tabular}{c}
\rotatebox{-90}{\includegraphics[height=10cm]{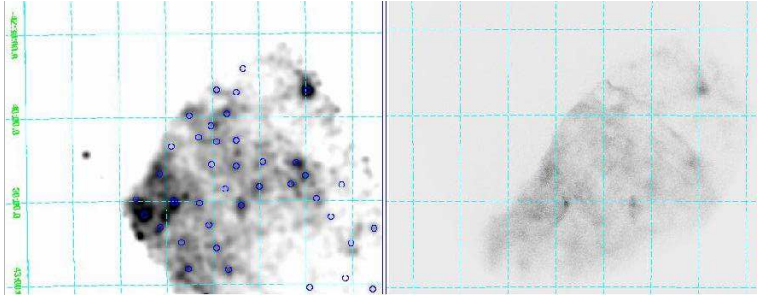}}
\end{tabular}
\end{center}
\caption[Puppis-A]
{ \label{fig:puppis}
A comparison of an adaptively smoothed, exposure corrected, XMM slew image of the Puppis-A SNR (left) with a ROSAT-HRI, 7000 second pointed exposure, taken in 1992 (right). The 
XMM image is limited by the 0.5 degree width of the pn camera while the ROSAT
image is cut by the bottom edge of the HRI detector. The small circles in the XMM image indicate the positions
of detections made by the source search software.}
\end{figure}

\begin{figure}
\begin{center}
\begin{tabular}{c}
\rotatebox{0}{\includegraphics[height=5cm]{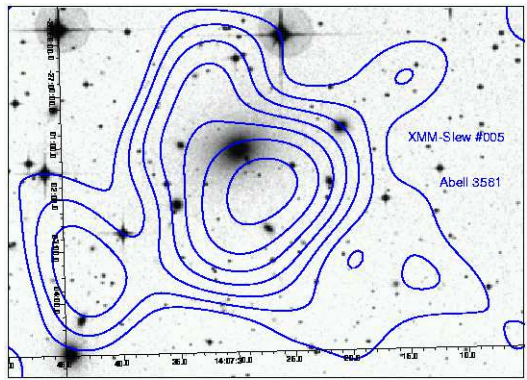}}
\end{tabular}
\end{center}
\caption[Abell 3581]
{ \label{fig:ab3581}
Contours of X-ray emission from a single slew across the
galaxy cluster Abell 3581, superimposed on a DSS
image.} 
\end{figure}

The good spatial resolution and low background of \xmm allows the
slew survey to usefully image bright extended sources. The very bright, large, SNR PUPPIS-A was slewed over in 2002 and the X-ray emission shows structure 
in a smoothed
image which correlates well with that seen in a pointed ROSAT HRI observation
 (Fig.~\ref{fig:puppis}).
Nearby clusters of galaxies, such as Abell 3581, can also be clearly 
detected as extended (Fig.~\ref{fig:ab3581}) and there is the possibility that at the
faint end of the survey new clusters or galaxy groups, too
small to be detected as extended in the RASS, will be discovered.

\section{Summary}

The \xmm slew data constitute a wide area (currently 20\% of the sky) shallow
survey whose soft band flux limits are sufficiently deep to provide an
interesting comparison with the RASS and whose hard band limits
represent an order of magnitude improvement over previous missions.
Several technical challenges have been overcome, particularly in understanding 
and refining the astrometry and in rejecting spurious sources. The
astrometry is good, with a 1 sigma position error of 8 arcseconds,
easily sufficient to allow an optical follow-up of these high flux
X-ray sources. Data processing is progressing well and the final 
total energy band catalogue should contain between three and five 
thousand sources,
depending on the final choice of maximum likelihood detection threshold
employed. The hard band catalogue will contain between 400 and 800
sources.

\acknowledgments     
 
This research has made use of the SIMBAD database,
operated at CDS, Strasbourg, France. We thank the Astrophysical Virtual Observatory (AVO) for providing software tools. AVO has been awarded financial support by the European Commission through contract HPRI-CT-2001-50030 under the 5th Framework Programme for research, technological development and demonstration activities. Based on data obtained with XMM-Newton, an ESA science mission with
instruments and contributions directly funded by ESA Member States and NASA.
Congratulations are due to the designers and operators of the
attitude control system for providing an accurate satellite 
pointing position throughout slewing manoeuvers.
We thank Georg Lammers and Herman Brunner for providing source
search software which proved to be robust on slew data.
We also thank Mark Tuttlebee, Pedro Rodriguez, John Hoar and Aitor Ibarra
for their help with understanding the slew attitude history files.




\bibliography{report}   
\bibliographystyle{spiebib}   

\end{document}